%% file: main.tex
\documentclass[aps,reprint,superscriptaddress,preprintnumbers]{revtex4-2}
\usepackage{amsmath}
\usepackage{xr-hyper}
\usepackage{hyperref}
\usepackage{isotope}
\usepackage{mhchem}
\usepackage{xcolor}
\usepackage{graphicx}
\usepackage{mathrsfs}
\usepackage{dcolumn}
\usepackage{bm}
\usepackage{cases}
\usepackage{multirow,booktabs}
\usepackage{makecell}
\usepackage{amsmath}
\usepackage{amssymb}
\usepackage{sidecap}
\usepackage{graphicx}
\usepackage{here}
\usepackage{braket}
\usepackage{adjustbox}

\usepackage{tikz}
\usepackage{float}
\usepackage{tabularx}
\usepackage{soul}
\usepackage[normalem]{ulem}
\hypersetup{
  breaklinks=true,
  colorlinks=true,
  linkcolor=blue,
  urlcolor=blue,
  citecolor=blue
}

\usepackage{xcolor}
\definecolor{pastelgray}{rgb}{0.81, 0.81, 0.77}
\definecolor{beaublue}{rgb}{0.9, 0.9, 0.93}

\usepackage{todonotes}

\begin{document}


\title{\textbf{Statistical properties of neutron-induced reaction cross sections using random-matrix approach} 
}%

\author{K.~Fujio}
\email{kazuki.fujio@lanl.gov}
\author{T.~Kawano}
\author{A.E.~Lovell}
\author{D.~Neudecker}
\affiliation{Los Alamos National Laboratory, Los Alamos, New Mexico 87545, USA
}%

\date{\today}

\begin{abstract}
We investigate the statistical properties of neutron-induced nuclear reactions on $^{238}$U using the GOE-$S$-matrix model, in which the Gaussian Orthogonal Ensemble (GOE) is embedded into the scattering ($S$) matrix. 
The GOE-$S$-matrix model does not require any experimental values of the average level spacing $D$ and average decay width $\Gamma$ with their statistical distributions, but the model is fully characterized by the channel transmission coefficients used in the Hauser-Feshbach theory.
We demonstrate that the obtained compound nucleus decay width distribution resembles the $\chi$-squared distribution with the degree of freedom greater than unity.
This approach enables us to generate fluctuating cross sections while preserving requisite unitarity and accounting for interference between resonances. 
By comparing the calculated cross section distribution with that from $R$-matrix theory, we demonstrate a smooth transition from the resolved resonance region to the continuum region.
\end{abstract}

\maketitle


\section{Introduction}
\input{introduction2}

\section{Theory to calculate fluctuating cross sections}
\input{theory}

\section{Results and discussion}
\input{results}

\section{Conclusion}
\input{conclusion}

\acknowledgments
We would like to thank N. Walton for fruitful discussions.
Los Alamos National Laboratory is operated by Triad National Security, LLC, for the National Nuclear Security Administration of the US Department of Energy under Contract No. 89233218CNA000001. Research reported in this publication was supported by the U.S. Department of Energy LDRD program at Los Alamos National Laboratory (Project No. 20240031DR).

\bibliographystyle{apsrev4-2}
\bibliography{references}

\end{document}

%% file: introduction2.tex
When a low-energy neutron interacts with a nucleus, sharp and well resolved resonances appear in the reaction cross sections due to the formation of states in the compound nucleus.
This energy range, typically extending up to a few keV depending on the nucleus, is often referred to as the resolved resonance region (RRR).
The RRR is often described using $R$-matrix theory~\cite{Lane1958}, which characterizes the resonances by a set of parameters such as the resonance energies and decay widths.
While the ratio of the average width to the average resonance spacing, $\langle\Gamma\rangle / D$, is small at very low energies, this ratio increases with the incident neutron energy.  
At higher energies, the ratio becomes greater than unity, and the individual resonances can no longer be distinguished from one another, resulting in less fluctuation in the observed cross sections.
This transition usually happens in the keV energy region for actinides. 
Beyond the transition region, only the energy-averaged cross section becomes meaningful.
The Hauser-Feshbach theory~\cite{Hauser1952} describes these smooth average cross sections, where the average decay width is replaced by the transmission coefficients.

The unresolved resonance region (URR) is the transition region from RRR to smooth cross sections, where both the fluctuating and average cross sections characterize the statistical properties of nuclear reactions.
Theoretical models for the RRR and higher energy region are well established, and the average cross sections in the URR can be calculated using the existing Hauser-Feshbach theory with proper width fluctuation corrections~\cite{Dresner1957}. 
Cross section fluctuations in the URR have been studied under simplified assumptions, such as the single-level Breit–Wigner (SLBW) formula~\cite{Breit1936} by assuming some statistical properties of resonances.
However, the SLBW formalism has several limitations: it can give negative cross sections due to broken unitarity, and it does not properly account for interference between resonances.  
Even a more sophisticated approach, such as employing the $R$ matrix with Reich-Moore approximation~\cite{Reich1958}, still needs to know the statistical distributions of resonances~\cite{Koyumdjieva2010,Holcomb2017}.
As a result, the statistical properties in the URR lack a solid theoretical foundation and remain largely unexplored.

Although no reliable theoretical model has yet been established for these statistical properties, a promising approach has been proposed based on the concepts of random-matrix theory, providing potential starting points for elucidating the statistical nature of nuclear reactions.
The idea of the random-matrix theory was first introduced into nuclear physics by Wigner, who hypothesized that the statistical properties of random matrices can describe those of resonance level spacings in nuclei~\cite{Wigner1951}.
The Gaussian Orthogonal Ensemble (GOE), which consists of real symmetric matrices whose elements are random numbers following a Gaussian distribution, has been employed to study the statistical properties of compound nuclei~\cite{Weidenmueller2009,Mitchell2010}.
The distribution of eigenvalue spacings in the GOE closely matches the experimentally observed distribution of resonance level spacings, known as the Wigner distribution~\cite{Wigner1951}. 
This remarkable agreement provides strong evidence supporting both the validity of Wigner’s hypothesis and the applicability of random matrix theory to the statistical description of nuclear resonances.
Together with an assumption of the $\chi$-squared distribution with 1 degree of freedom (the Porter-Thomas distribution~\cite{Porter1956}) that is observed in the decay width, average cross sections can be studied~\cite{Brown2018,Nobre2023}.

Verbaarschot, Weidenm\"uller, and Zirnbauer~\cite{Verbaarschot1985} incorporated the GOE into the scattering ($S$) matrix, and derived an analytical expression of the proper average, where three-fold integration is involved.
In order to distinguish their approach from the GOE itself, we refer to the $S$ matrix including the GOE as the GOE-$S$-matrix model.
While a GOE-inspired statistical theory where the Wigner and Porter-Thomas distributions are assumed, the GOE-$S$-matrix is characterized by
channel transmission coefficients only.

The GOE-$S$-matrix model enables us to estimate the width fluctuation correction factor~\cite{Moldauer1975a} that is necessary to calculate the average cross section in more realistic cases (note that the GOE-$S$-matrix model itself cannot be applied to practical cross section calculations so far).
Kawano, Talou, and Weidenm\"uller~\cite{Kawano2015} investigated the statistical properties of $S$ matrices and cross sections by using the GOE-$S$-matrix model with a Monte Carlo technique, which is equivalent to solving the three-fold integration.
This model also has been employed in various nuclear reaction processes~\cite{Bertsch2018,Fanto2018,Bertsch2024,Weidenmuller2024}.
In contrast to the GOE itself, the GOE-$S$-matrix model provides information on all the statistical properties of the average cross sections and their widths in a consistent framework.
It is not yet clear whether the resonance width distribution of the GOE-$S$-matrix model would follow the Porter–Thomas distribution or not.

The objectives of this study are to elucidate the statistical properties of nuclear reactions by employing the GOE-$S$-matrix model in a realistic case.
We explore a smooth and natural transition from the RRR to the Hauser-Feshbach region by providing consistent transmission coefficients across these regions.  We calculate the energy-average $S$-matrix elements from the $R$-matrix parameters in the RRR.  
These average matrix elements will be connected to the optical model $S$ matrix, which enables us to naturally incorporate the energy dependence of the transmission coefficients, as well as the contribution of higher orbital angular momenta.  
Once the energy-average $S$-matrix elements (or transmission coefficients) are fixed, the GOE-$S$-matrix model uniquely provides the average cross section as well as its statistical properties.
One of the advantages of this method is that it no longer requires the use of experimental $\langle\Gamma\rangle$ and $D$, but the GOE-$S$-matrix model is controlled by the transmission coefficients only.
This implies that we obtain the identical average cross sections when the same transmission coefficients are used in the Hauser-Feshbach model calculation with width fluctuation correction.
Using $^{238}$U as the target nucleus, we compare the statistical properties of the compound nucleus reaction given by the GOE-$S$-matrix model with the SLBW approach.  
The properties include average cross sections and their distributions, and the distributions of decay widths.

%% file: theory.tex
\subsection{GOE-$S$-matrix model}
In the GOE-$S$-matrix approach, where the GOE Hamiltonian is used to calculate the $S$ matrix, fluctuating cross sections can be calculated using parameters~\cite{Verbaarschot1985,Kawano2015}, such as the number of levels $n$, the number of channels $m$, and transmission coefficients for each channel.
The GOE Hamiltonian $H^{\rm (GOE)}_{\mu\nu}$ is a time-reversal invariant and represented by an $n\times n$ real symmetric matrix whose elements are random numbers following a Gaussian distribution.
The mean value is 0, and the second moment is given by
\begin{eqnarray}
    \overline{H^{\rm (GOE)}_{\mu\nu}H^{\rm (GOE)}_{\rho\sigma}}=
    \frac{\lambda^2}{n}(\delta_{\mu\rho}\delta_{\nu\sigma}+\delta_{\mu\sigma}\delta_{\nu\rho}),
\end{eqnarray}
where $\lambda$ is a scaling factor that affects the level spacing, and all $\delta$'s are the Kronecker deltas.
The indices $\mu$, $\nu$, $\rho$, and $\sigma$ range from 1 to $n$.
The energy in the eigenchannel $E_{\lambda}$ expressed in units of $\lambda$ is defined in the GOE-$S$-matrix model.
The eigenvalues of GOE are distributed within the range $E_{\lambda}=-2\lambda$ to $2\lambda$, and their density forms the famous semi-circle distribution.
We introduce the hard-sphere phase shift $\phi = k r$, where $k$ is the wave number and $r$ is the channel radius, based on an analogy with the collision matrix expression.
The $S$ matrix for a reaction from channel $a$ to $b$ is written in
\begin{eqnarray}
\label{eq:Smat}
    S^{\rm (GOE)}_{ab}&=&e^{-i\left(\phi_a+\phi_b\right)}\left\{\delta_{ab}-2i\pi\sum_{\mu\nu}W_{a\mu}\left(D^{-1}\right)_{\mu\nu}W_{\nu b}\right\}. \nonumber \\
\end{eqnarray}
We determine $r_a$ to reproduce the experimental total cross section.
The radius $r_b$ disappears in our model when we take $|S^{\rm (GOE)}_{ab}|^2$ to calculate the inelastic scattering cross section.
The denominator $D_{\mu\nu}$ is an $n\times n$ matrix analogous to the level matrix that arises in the formulation of the collision matrix:
\begin{eqnarray}
\label{eq:denom}
    D_{\mu\nu}=E_{\lambda}\delta_{\mu\nu}-H^{\rm (GOE)}_{\mu\nu}+i\pi\sum_cW_{\mu c}W_{c\nu}.
\end{eqnarray}
Since the eigenvalues correspond to the experimentally observed resonances, we need to map $E_{\lambda}$ onto the laboratory energy. 
In this study, the actual incident neutron energy is used for $k_a$, and $E_\lambda=0$ (at the center of semi-circle) is taken to be the corresponding incident neutron energy.
The quantity $W_{a\mu}$ represents an element of the $n\times m$ coupling matrix $W$.
The matrix $W$ is calculated by the coupling strength matrix $G$ and the orthogonal matrix $\mathcal{O}$ which is obtained from the diagonalization of $G^{T}G$:
\begin{eqnarray}
    W=G\mathcal{O}.
\end{eqnarray}
The matrix $G$ is constructed by the coupling vector elements $w_a$:
\begin{eqnarray}
    w_{a}&=&\sqrt{\frac{x_a}{\pi}}, \\
    x_{a}&=&\frac{2}{t_a}(1-\sqrt{1-t_a})-1,
\end{eqnarray}
where $t_a$ is the input transmission coefficient for channel $a$.

In our calculation for $^{238}$U cross sections, we employ the generalized transmission coefficients~\cite{Satchler1963,Kawano2009,Kawano2016} calculated in CoH$_3$~\cite{Kawano2021} using an optical model potential parameter set suggested by Soukhovitskii, \textit{et al}~\cite{Soukhovitskii2004}.
We confirm that the $S$ matrix calculated using this optical model potential reproduces the energy-averaged $R$ matrix $\braket{R(E)}$, which is calculated by
\begin{eqnarray}
    \braket{R(E)}=R(E+iI),
\end{eqnarray}
where $I$ is a Lorentzian width, and $R(E)$ is the $R$ matrix calculated by the Reich-Moore resonance parameters in ENDF/B-VIII.0.
This indicates that the resulting transmission coefficients are consistent between the RRR and the continuum region.
For neutron capture channels, the transmission coefficients are calculated by using the giant dipole resonance model: the Kopecky-Uhl~\cite{Kopecky1990} model for E1 transitions, and the Brink-Axel model~\cite{Brink1957,Axel1962} for the higher multi-polarities of M1 to E3 transitions. 
The scissors mode~\cite{Mumpower2017} is also included in M1 transition.
The Gilbert-Cameron level density formula~\cite{Gilbert1965} is employed, using a level density parameter~\cite{Kawano2006} whose shell correction and pairing energy are determined based on the KTUY05 mass formula~\cite{Koura2005}.
The level density parameter is adjusted to reproduce the average level spacing of 20.3 eV of $^{239}$U~\cite{Capote2009}.
Discrete level data are taken from RIPL-3~\cite{Capote2009}.
$\gamma$-ray transitions include both transitions between continuum states and transitions from continuum states to discrete levels.
The number of neutron capture channels leaves somewhat arbitrary due to the binning of the continuum state.
This may cause discrepancies between the calculated average cross sections and those obtained from the Hauser-Feshbach theory when the number of capture channels is not large enough.
To mitigate this dependence, the $\gamma$-ray transitions are divided into 10 channels in this study.
The number of the inelastic channels are determined based on the corresponding excitation levels as well as possible number of neutron partial waves.
The fission channel is neglected, as the fission cross section is negligible in the case of $^{238}$U below 1 MeV.
The number of levels is set to $n=100$ in this study.
In fact $n$ is not a parameter but it should be large enough in our model. 
We confirmed that $n = 100$ satisfies this condition because the calculation results converge for $n\geq100$.

The cross section for a reaction from channel $a$ to $b$ is calculated by
\begin{eqnarray}
\label{eq:cs}
    \sigma_{ab}=\frac{\pi}{k_a^2}g_J\left|\delta_{ab}-S^{\rm (GOE)}_{ab}\right|^2,
\end{eqnarray}
where $g_J$ is the spin factor.

\subsection{Decay amplitude}
The experimentally observed decay width is represented by the ensemble average of the square of the reduced width amplitude in the $R$-matrix formalism.
We reformulate Eq.~(\ref{eq:Smat}) in the form of $R$-matrix formalism to investigate the distribution of the decay width in the random-matrix approach.
The $R$ matrix can be expressed in terms of the reduced width amplitude $\gamma_{a\sigma}$ as follows:
\begin{eqnarray}
    R_{ab}(E)=\sum_{\sigma}\frac{\gamma_{a\sigma}\gamma_{\sigma b}}{E_{\sigma}-E}.
\end{eqnarray}
The $K$ matrix is related to the $S$ matrix by:
\begin{eqnarray}
    S^{\rm (GOE)}_{ab}=\left(\frac{1-iK^{\rm (GOE)}}{1+iK^{\rm (GOE)}}\right)_{ab},
\end{eqnarray}
where
\begin{eqnarray}
    K^{\rm (GOE)}_{ab}(E)&=&\sum_{\sigma}\frac{\tilde{W}_{a\sigma}\tilde{W}_{\sigma b}}{E-E_{\sigma}}.
\end{eqnarray}
Here, the amplitude $\tilde{W}_{a\sigma}$ is calculated by $W$ and the orthogonal matrix $\mathcal{O}$.
The matrix $\mathcal{O}$ is obtained from the diagonalization of $H^{\rm (GOE)}$, with its eigenvalues denoted by $E_{\sigma}$:
\begin{eqnarray}
\label{eq:tildeW}
    \tilde{W}_{\sigma a}=\sqrt{\pi}\sum_{\nu}\mathcal{O}_{\sigma\nu}W_{\nu a}, \\
    \mathcal{O}^{-1}H^{\rm (GOE)}\mathcal{O}={\rm diag}(E_\sigma).
\end{eqnarray}
The amplitude $\tilde{W}_{a\sigma}$ is related to $\gamma_{a\sigma}$ in the $R$-matrix formalism by the relation $\tilde{W}_{a\sigma}=\sqrt{\pi}\gamma_{a\sigma}$.

Resonances are represented as poles of the $S$ matrix.
The $S$ matrix can be written in the form of a pole expansion as follows:
\begin{eqnarray}
    S^{\rm (GOE)}_{ab}(E)&=&\delta_{ab}-i\sum_{\sigma}\frac{\tilde{W}_{a\sigma}\tilde{W}_{\sigma b}}{E-E_{\sigma}}, 
\end{eqnarray}
where the amplitude $\tilde{W}_{a\sigma}$ is calculated with $W$ and the orthogonal matrix $\mathcal{O}$ in the same manner as in Eq.~(\ref{eq:tildeW}).
The matrix $\mathcal{O}$ is obtained from the diagonalization of $H^{\rm (GOE)}-i\pi W^TW$, with its eigenvalues denoted by $E_{\sigma}$:
\begin{eqnarray}
    \mathcal{O}^{-1}(H^{\rm (GOE)}-i\pi W^TW)\mathcal{O}={\rm diag}(E_\sigma).
\end{eqnarray}

\subsection{Single-level Breit-Wigner formula}
Evaluated nuclear data libraries, such as ENDF/B-VIII.0~\cite{Brown2018_endf}, JEFF-3.3~\cite{Plompen2020}, and JENDL-5~\cite{Iwamoto2023}, provide average resonance parameters in the URR for each orbital angular momentum $l$ and total angular momentum $J$.
The average resonance parameters include the average level spacing $D$, average widths for the reduced neutron $\braket{\Gamma_n}$, radiative neutron capture $\braket{\Gamma_\gamma}$, fission $\braket{\Gamma_f}$, and other competing reactions, such as inelastic scattering $\braket{\Gamma_{n'}}$.
It remains experimentally unclear whether the cross section exhibits fluctuations due to the finite energy resolution of measurement, and if so, what the magnitude of those fluctuations is.
By assuming statistical distributions for $D$ and $\braket{\Gamma}$'s, one can replicate the fluctuating cross sections, whose average properties coincide with the provided values of $D$ and $\braket{\Gamma}$'s even if the simulated fluctuations do not exactly match experimental observations. 
The Wigner distribution $P_{\rm W}$ is often employed for $D$:
\begin{eqnarray}
\label{eq:W}
    P_{\rm W}(x) = \frac{\pi}{2}x\exp{\left(-\frac{\pi x^2}{4}\right)}, 
\end{eqnarray}
where $x = d/D$ represents the eigenvalue spacing normalized by the average spacing $D$.
The Porter-Thomas distribution $P_{\rm PT}$ is used for $\braket{\Gamma}$'s:
\begin{eqnarray}
\label{eq:PT}
    P_{\rm PT}(x)=\frac{1}{\sqrt{2\pi x}}\exp{\left(-\frac{x}{2}\right)},
\end{eqnarray}
where $x = \Gamma/\braket{\Gamma}$ represents the normalized width.
In this work, $P_{\rm PT}$ is applied to the widths of elastic and inelastic scattering channels.
By using generated energies and decay widths for individual resonances according to Eqs.~(\ref{eq:W}) and~(\ref{eq:PT}), the elastic cross section $\sigma_{aa}$ and a reaction cross section from channel $a$ to $b$, $\sigma_{ab}$, are reconstructed by~\cite{ENDFmanual}
\begin{widetext}
\begin{eqnarray}
\label{eq:SLBWCS}
    \sigma_{aa}(E)&=&\sum_l\left\{(2l+1)\frac{4\pi}{k_a^2}\sin^2\phi_l
    +\frac{\pi}{k_a^2}\sum_Jg_J\sum_{r}\frac{\Gamma^{l\,2}_{n,r}-2\Gamma^{l}_{n,r}\Gamma^{l}_{tot,r}\sin^2\phi_l+2(E-E'_r)\Gamma^{l}_{n,r}\sin(2\phi_l)}{(E-E_r')^2+\frac{1}{4}\Gamma^{l\,2}_{tot,r}}\right\}, \\
    \sigma_{ab}(E)&=&\sum_{l}\left\{\frac{\pi}{k_a^2}\sum_Jg_J\sum_r\frac{\Gamma^{l}_{n,r}\Gamma^{l}_{b,r}}{(E-E'_r)^2+\frac{1}{4}\Gamma^{l\,2}_{tot,r}}\right\},
\end{eqnarray}
\end{widetext}
where $E_r$ represents the energy of the $r$-th resonance, and $\Gamma^l_{tot,r}$ denotes the total width at the $r$-th resonance.
The total width is defined as the sum of the neutron width $\Gamma^l_{n,r}$, radiative capture width $\Gamma^l_{\gamma,r}$, and inelastic scattering width $\Gamma^l_{{n'},r}$ if the fission channel is negligible:
\begin{eqnarray}
\label{eq:tot_width}
    \Gamma^l_{tot,r}=\Gamma^l_{n,r}+\Gamma^l_{\gamma,r}+\Gamma^l_{{n'},r}.
\end{eqnarray}
The quantity $\phi_l$ in Eq.~(\ref{eq:SLBWCS}) is hard-sphere phase shift and is calculated with the channel radius provided in the evaluated data. 

%% file: results.tex
\subsection{Statistical property of width distributions}
\label{subsec:statistical}
Since the GOE-$S$-matrix model does not assume any statistical distributions for the decay width, it is not obvious what kind of decay width distribution is finally formed.
We investigate the width distributions for both elastic and neutron capture channels using the $K$-matrix representation and the pole expansion of the $S$ matrix, respectively, based on 0.2 million GOE samples.
The transmission coefficient reproducing the average cross section is relatively small in the URR of $^{238}$U; that is, the coupling between outgoing and incoming channels is weak.
In this case, the $K$-matrix representation and the pole expansion of the $S$ matrix exhibit approximately equivalent results.
Therefore, we present only the distributions obtained from the $K$-matrix representation.

\begin{figure}[H]
    \centering
    \includegraphics[width=\linewidth]{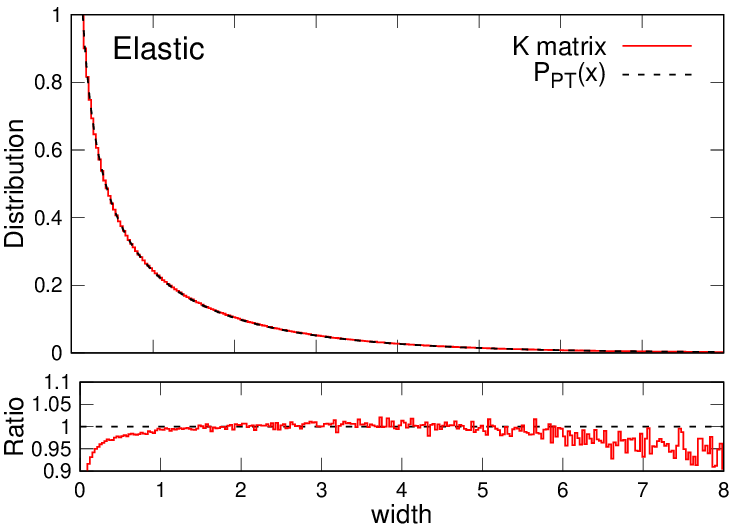}
    \caption{Width distribution of the elastic channel at an incident neutron energy of 0.02 MeV for the $s$-wave (top panel) and the ratio to the Porter–Thomas distribution (bottom panel).}
    \label{fig:width_0.02MeV_elastic}
\end{figure}

\begin{figure}[H]
    \centering
    \includegraphics[width=\linewidth]{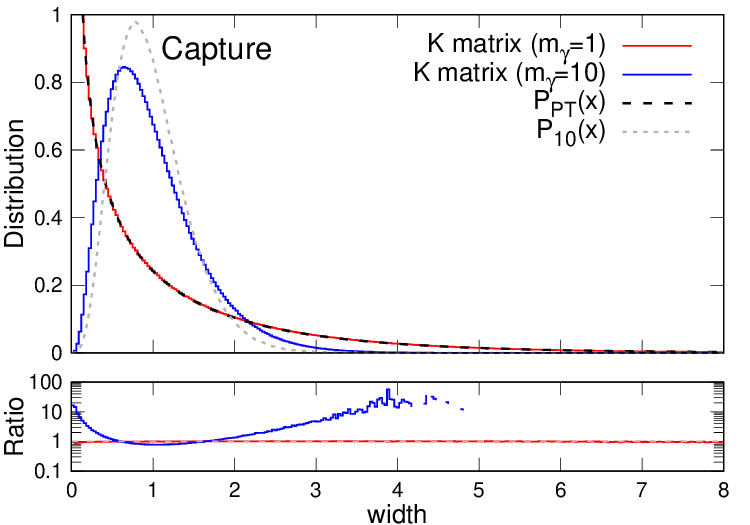}
    \caption{Width distribution of neutron capture channel at an incident neutron energy of 0.02 MeV for the $s$-wave (top panel) and the ratio to the Porter–Thomas distribution (bottom panel). The gray dashed line corresponds to a distribution obtained by summing 10 independent $\chi$-squared variables, namely, the $\chi$-squared distribution with the 10 degree of freedom.}
    \label{fig:width_0.02MeV_radiative}
\end{figure}

Figures~\ref{fig:width_0.02MeV_elastic} and \ref{fig:width_0.02MeV_radiative} show the width distributions for the elastic and neutron capture channels at an incident neutron energy of 0.02 MeV for the $s$-wave. 
The bottom panel of Fig~\ref{fig:width_0.02MeV_elastic} presents the ratio of the calculated width distributions to the Porter–Thomas distribution of Eq.~(\ref{eq:PT}).
For the neutron channel, the distribution obtained from the $K$-matrix representation almost coincides with the Porter-Thomas distribution.
The distribution deviates from the Porter–Thomas distribution in the vicinity of 0 and in the tail region. 
The degree of freedom $\nu$ of the calculated distribution is determined by the least-squares method using the $\chi$-squared distribution $\chi^2_{\nu}$:
\begin{eqnarray}
    \chi^2_{\nu}(x,\nu)=\frac{1}{2^{\frac{\nu}{2}}\Gamma\left(\frac{\nu}{2}\right)}x^{\frac{\nu}{2}-1}\exp{\left(-\frac{x}{2}\right)},
\end{eqnarray}
where $\Gamma(\nu/2)$ is the Gamma function.
The obtained $\nu$ is 1.14 in both the $K$-matrix representation and the pole expansion of the $S$ matrix.
This result, namely that $\nu$ is greater than 1, is consistent with the prescriptions of Moldauer~\cite{Moldauer1980}, and of Kawano and Talou~\cite{Kawano2015}.

When the neutron capture is treated as a single decay channel ($m_\gamma=1$), the resulting width distribution also becomes similar to the Porter-Thomas distribution, as shown by the red line in Fig.~\ref{fig:width_0.02MeV_radiative}. 
In contrast, when it is described by multiple independent channels ($m_\gamma=10$), the resulting distribution deviates from the Porter-Thomas form, as illustrated by the blue line in Fig.~\ref{fig:width_0.02MeV_radiative}. 
The distribution exhibits a shape more consistent with a $\chi$-squared distribution with more than 1 degree of freedom.
Let $z_{i,{\rm PT}}$ denote independent random variables sampled from the Porter-Thomas distribution.
The sum of these $m_\gamma$ independent variables is defined as
\begin{eqnarray}
    Z_{m_\gamma}(x)=\sum_{i=1}^{m_\gamma}w_iz_{{\rm PT},i}(x),
\end{eqnarray}
where $w_i$ denotes the weights assigned to each neutron capture channel.
The sum of $w_i$ is determined to be 1, and all weights are assumed to be equal.
The probability density function of $Z_{10}(x)$ is denoted by 
$P_{10}(x)$, which is shown by the gray dashed line in Fig.~\ref{fig:width_0.02MeV_radiative}.
The bottom panel of the figure shows the ratios of the calculated width distributions to $P_{\rm PT}$ and $P_{10}$, respectively.
As $m_\gamma\rightarrow\infty$, the distribution $P_{m_\gamma}(x)$ approaches a Gaussian distribution with a mean of 1, in accordance with the central limit theorem. 
Similarly, the mean of the distribution calculated by the GOE-$S$-matrix model is considered to converge to 1 as $m_\gamma$ increases.
These tendencies indicate that the Porter-Thomas distribution for the neutron capture channel is not satisfied when many of the $\gamma$-decay channels are lumped.
The neutron capture channel includes enormous number of final states after $\gamma$-ray emission due to the high level density of the compound nucleus. 
Since including all the final states is impractical, our modeling represents these states by 10 effective channels. 
This approach is more realistic than lumping all $\gamma$-ray transitions into one channel because spin and parity conservations are properly incorporated into the $\gamma$-decay.
For small values of $m_\gamma$, differences in the distribution by the GOE-$S$-matrix model and $P_{10}(x)$ arise from the unitarity condition imposed on the $S$ matrix.

Since the distribution of the decay width is derived through an orthogonal transformation of Eq.~(\ref{eq:tildeW}), no specific statistical distribution is assumed for the decay width in this work.
Note that there is no stochasticity in the coupling matrix $W$ because it is constructed by the transmission coefficients only. 
The width distribution appears because of GOE embedded in the energy denominator of Eq.~(\ref{eq:denom}).
The degrees of freedom $\nu$ of the $\chi$-squared distribution for both the elastic and neutron capture channels consistently exceed 1. 
Similar behavior is observed across different incident neutron energies and partial waves.


\subsection{Fluctuating and average cross sections}
\begin{figure*}[htbp]
    \centering
    \includegraphics[width=\textwidth]{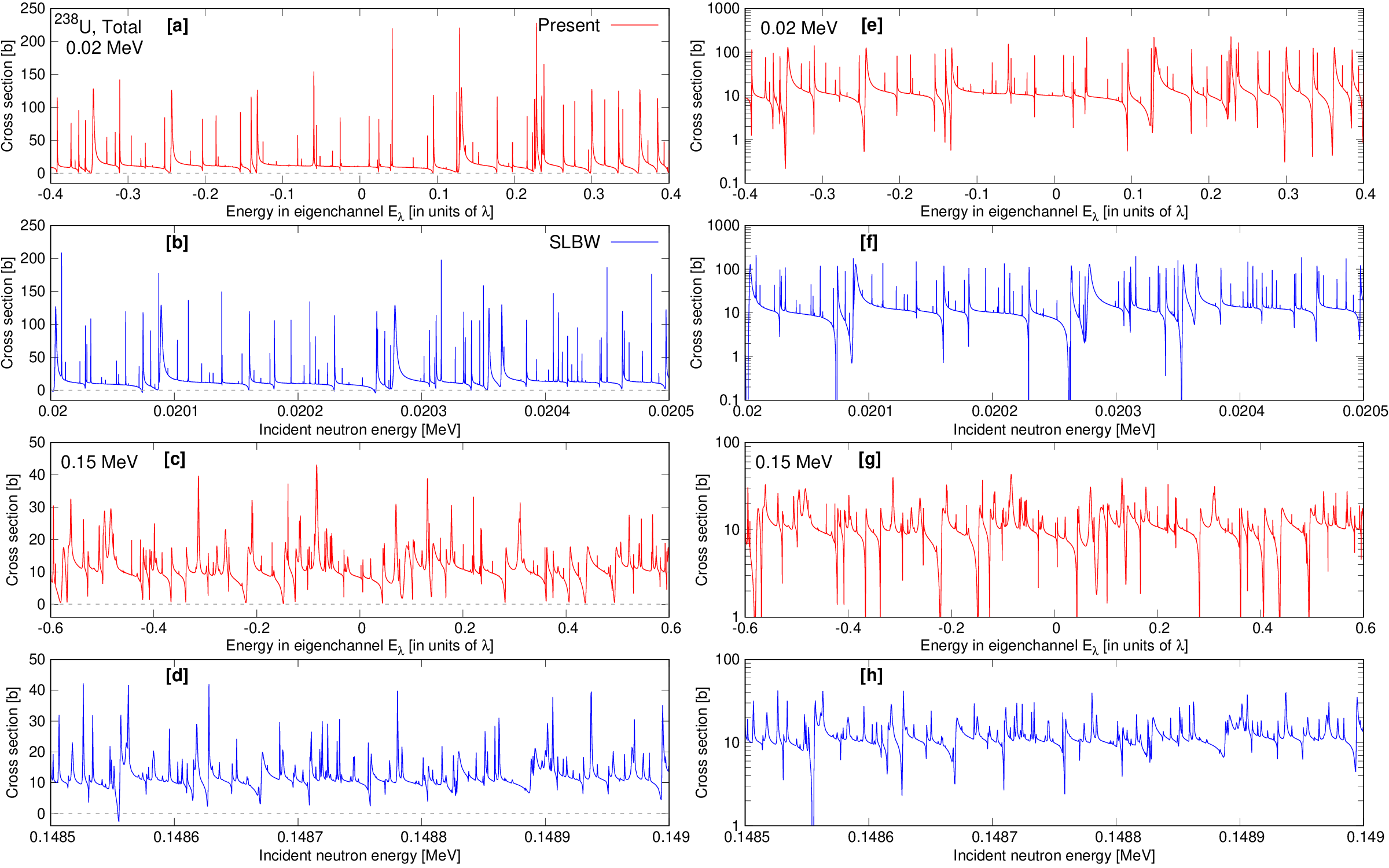}
  \caption{Total cross sections calculated by the GOE-$S$-matrix model ([a] and [c]) and those from SLBW ([b] and [d]) in the vicinity of incident neutron energies of 0.02 and 0.15 MeV. The right figures are the logarithmic-scale plots of the left figures.}
  \label{fig:resonance}
\end{figure*}

Fluctuating and average cross sections are calculated using the $S$ matrix obtained from the GOE-$S$-matrix model and are compared with those reconstructed by the SLBW formula using average resonance parameters taken from ENDF/B-VIII.0.
For the case of $^{238}$U, the average resonance parameters are provided for incident neutron energies between 0.02 and 0.15 MeV.
Figure~\ref{fig:resonance} shows the examples of the fluctuating total cross sections for incident neutron energies near 0.02 and 0.15 MeV. 
The GOE-$S$-matrix model produces the resonance structures similar to those obtained with the SLBW formula in the vicinity of both energies.
However, it tends to exhibit stronger interference effects between resonances than the SLBW, as clearly seen in the logarithmic-scale plots, particularly near the incident neutron energy of 0.15 MeV.
Note that $E_{\lambda}=0$ corresponds to the actual energies of 0.02 MeV (see sub-figure~[a] and [e]) and 0.15 MeV (see subfigure~[c] and [g]).
It is known that SLBW sometimes gives a negative cross section (broken unitarity), as seen some places in Fig.~\ref{fig:resonance}~[b] and [d]. 
The GOE-$S$-matrix model ensures the reconstructed cross sections are always positive since unitarity is always guaranteed.

\begin{figure}[H]
    \centering
    \includegraphics[width=\linewidth]{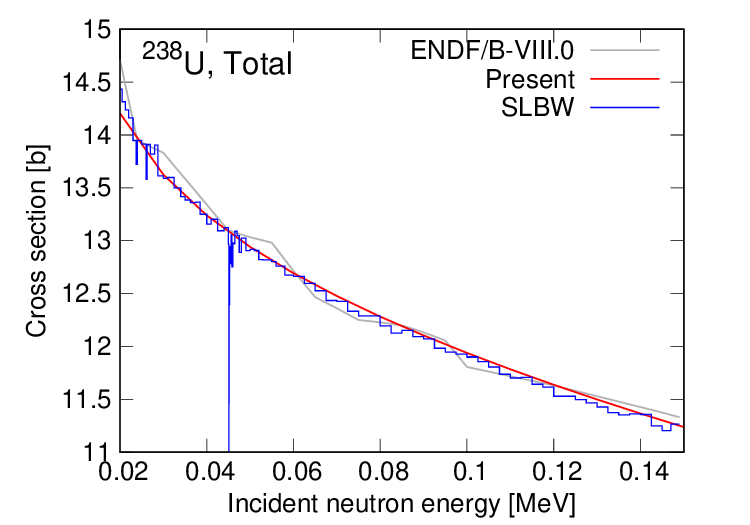}
    \caption{Calculated average total cross sections as a function of incident neutron energy.}
    \label{fig:average_crosssection_total}
\end{figure}

\begin{figure}[H]
    \centering
    \includegraphics[width=\linewidth]{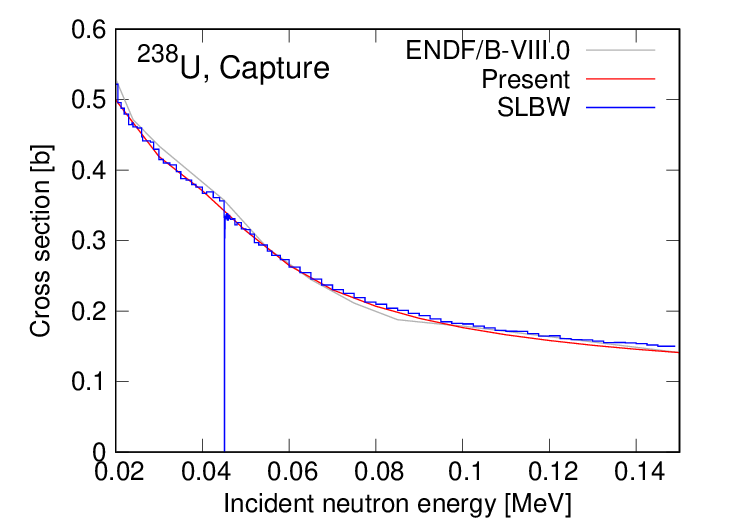}
    \caption{Calculated average neutron capture cross sections as a function of incident neutron energy.}
    \label{fig:average_crosssection_capture}
\end{figure}

By performing a large number of realizations of cross sections based on the Monte Carlo random sampling, we can calculate the average total and neutron capture cross sections as a function of incident neutron energy, which are shown in Figs.~\ref{fig:average_crosssection_total} and~\ref{fig:average_crosssection_capture}.
The cross sections from the GOE-$S$-matrix model are obtained by averaging 20 million events, whereas 50 realizations are used to calculate the average for the SLBW case.
A sudden unphysical dip seen near 0.04 MeV in the SLBW case is due to the average resonance parameters given in ENDF/B-VIII.0, where a denser incident energy grid than the average spacing is given near the inelastic scattering threshold energy.
The fluctuating cross sections obtained using the GOE-$S$-matrix model closely resemble those reconstructed with the SLBW formula, and the corresponding average cross sections are very similar to the SLBW average.
It would be worth to note that the URR parameters in ENDF/B-VIII.0 are obtained by fitting these parameters to experimental data in an energy dependent manner. 
Therefore, the choice of energy grid is rather arbitrary. 
The GOE-$S$-matrix model does not require the energy dependent $\braket{\Gamma}$'s and $D$, but the energy dependence is naturally introduced by the transmission coefficients.
In fact, we obtain the identical average cross sections when these transmission coefficients are used in the Hauser-Feshbach calculation with width fluctuation correction.

\subsection{Cross section distributions}
For more detailed comparisons, distributions of the cross sections are calculated by averaging 100 fluctuating samples generated by the GOE-$S$-matrix model and the SLBW formula.
The variances $V_{\sigma}$ of the distributions are calculated as functions of the incident neutron energy:
\begin{eqnarray}
    V_{\sigma}=\frac{1}{E_{\rm max}-E_{\rm min}}\int_{E_{\rm min}}^{E_{\rm max}}\left\{\sigma(E)-\bar{\sigma}\right\}^2dE,
\end{eqnarray}
where $\bar{\sigma}$ is the mean value for each energy bin given by
\begin{eqnarray}
    \bar{\sigma}=\frac{1}{E_{\rm max}-E_{\rm min}}\int_{E_{\rm min}}^{E_{\rm max}}\sigma(E)dE.
\end{eqnarray}
Both $\overline{\sigma}$ and $V_\sigma$ are normalized by the factor $\pi/k^2$ to make them dimensionless.
In the case of the GOE-$S$-matrix model, since $E_{\rm max}$ and $E_{\rm min}$ are taken at $E_{\lambda}=2\lambda$ and $E_{\lambda}=-2\lambda$, respectively, the average is independent of incident energy binning size. 
In the case of SLBW, the energy boundary for averaging is taken from the evaluated data, which may cause some energy binning dependence when the energy grid is not equidistant.

\begin{figure}[H]
    \centering
    \includegraphics[width=\linewidth]{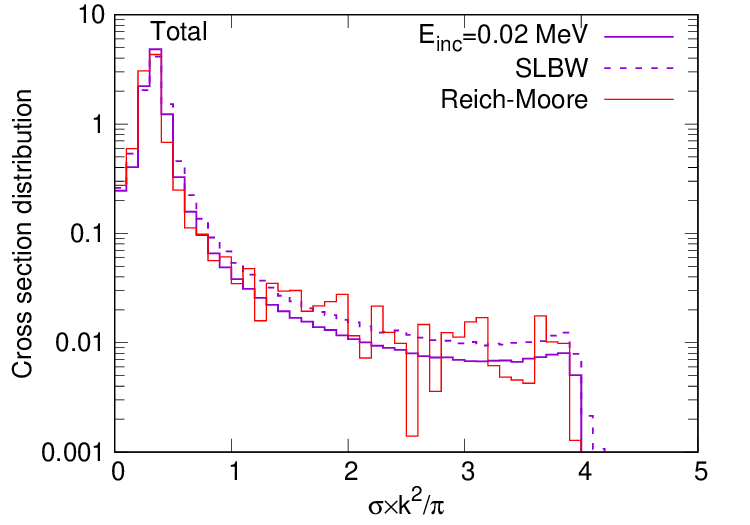}
    \caption{Distributions of total cross section normalized by $\pi/k^2$ at an incident neutron energy of 0.02 MeV.}
    \label{fig:distribution_total_0.02MeV}
\end{figure}

\begin{figure}[H]
    \centering
    \includegraphics[width=\linewidth]{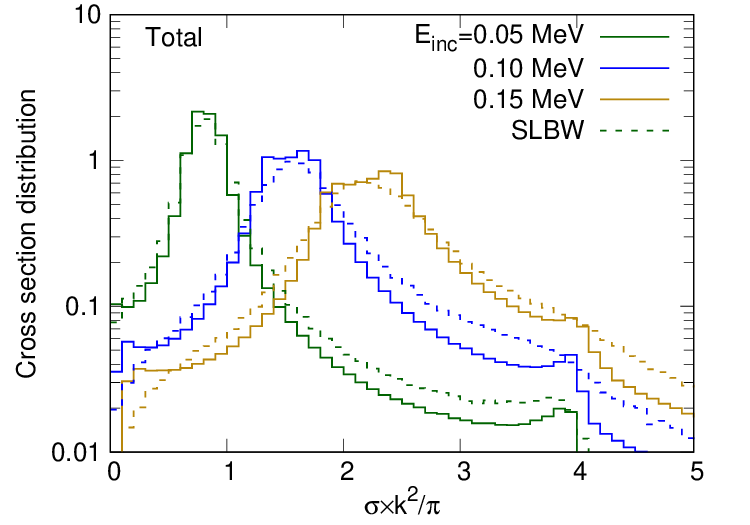}
    \caption{Distributions of total cross section normalized by $\pi/k^2$ at the incident neutron energies of 0.05, 0.10, and 0.15 MeV.}
    \label{fig:distribution_total}
\end{figure}

\begin{figure}[H]
    \centering
    \includegraphics[width=\linewidth]{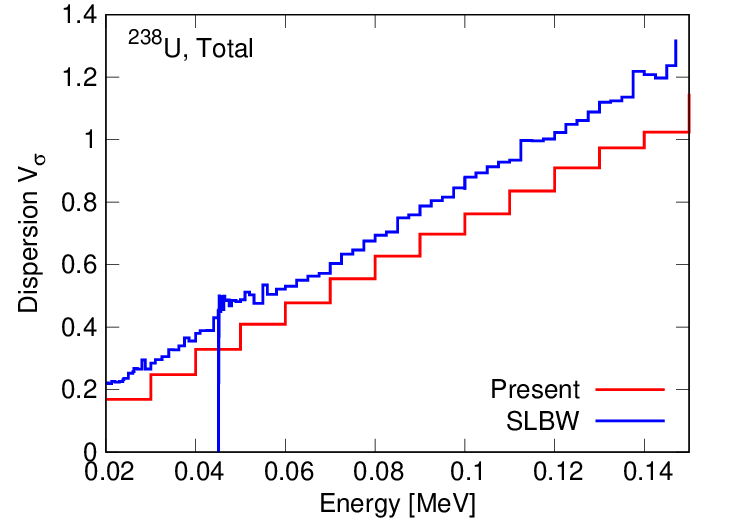}
    \caption{The dispersions $V_{\sigma}$ of the total cross section distributions.}
    \label{fig:dispersion_total}
\end{figure}

Figures~\ref{fig:distribution_total_0.02MeV} and ~\ref{fig:distribution_total} show the distributions of the total cross section normalized by the factor $\pi/k^2$, corresponding to incident neutron energies of 0.02, 0.05, 0.10, and 0.15 MeV, and Fig.~\ref{fig:dispersion_total} presents $V_{\sigma}$ as a function of the incident energy.
The cross section in the RRR is reconstructed by the Reich–Moore formalism~\cite{Reich1958} using resonance parameters, and the distribution represented by the red line in Fig.~\ref{fig:distribution_total_0.02MeV} is calculated based on this cross section.
Since 0.02 MeV is the upper limit of the RRR in the evaluated data, the distribution by the Reich–Moore formalism is applied only at 0.02 MeV.
The distribution from the GOE-$S$-matrix model is consistent with one from Reich-Moore, which indicates that the GOE-$S$-matrix model has the same statistical properties as the $R$-matrix theory in RRR, although the neutron capture channel is eliminated in the Reich-Moore approximation. 
It ensures smooth transition from RRR to URR.

The distributions obtained from both the GOE-$S$-matrix model and SLBW resemble a Lorentzian shape.
As the incident neutron energy increases, the peaks of the cross section distributions ($\bar{\sigma}$) shift toward higher values.
This shift is attributed to the increasing contribution of potential scattering and its interference with resonance scattering.
In addition, the widths of the distributions ($V_{\sigma}$) also increase.
This is because $\braket{\Gamma}/D$ increases with the incident neutron energy, which results in an apparent increase in width.
In the GOE-$S$-matrix model, the cross section distributions are primarily observed in the range $\sigma\times k^2/\pi=0$ to 4.
This is due to the unitarity of the $S$ matrix for $l=0$, which is the partial wave that mainly contributes to the cross section.
In this energy range, the tail of the distribution $\sigma\times k^2/\pi>4$ arises from contributions of partial waves with higher $l$ values due to the statistical factor $g_J$ in Eq.~(\ref{eq:cs}).
While the distribution obtained from the GOE-$S$-matrix model is narrower than that from SLBW, the GOE-$S$-matrix model shows higher probabilities to produce very small cross sections near $\sigma\times k^2/\pi=0$.
This is due to the resonance interference that produces deep valleys as seen in Fig.~\ref{fig:resonance}~[e] and [g], which is properly taken into account in the GOE-$S$-matrix model. 
Although SLBW includes the interference term in Eq.~(\ref{eq:SLBWCS}), the other channels---neutron capture and inelastic scattering---are just the sum of many level Breit-Winger that tends to fill the valleys.

\begin{figure}[H]
    \centering
    \includegraphics[width=\linewidth]{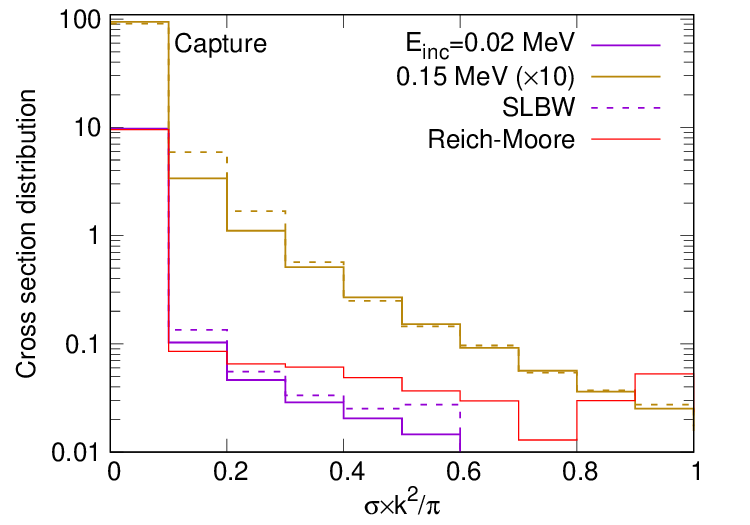}
    \caption{Distributions of neutron capture cross section normalized by $\pi/k^2$.}
    \label{fig:distribution_capture}
\end{figure}

Figure~\ref{fig:distribution_capture} shows the distributions of the neutron capture cross section normalized by $\pi/k^2$ at 0.02 and 0.15 MeV.
Since the contribution from the $s$-wave dominates, the cross sections are distributed in the range $\sigma\times k^2/\pi=0$ to 1 due to the unitarity of the $S$ matrix. 
Although the distribution obtained from the GOE-$S$-matrix model is similar to those derived from the SLBW and Reich–Moore formalisms at an incident neutron energy of 0.02 MeV, differences are observed in the tail region.
Both the GOE-$S$-matrix model and the SLBW show the same trend in which the width of the distributions increases with incident energy, yet slight deviations remain.
These differences primarily arise from the treatment of the neutron capture channel.
Both the SLBW and Reich-Moore formalisms do not preserve unitarity.
The $\gamma$-ray decay channels are lumped into a single channel in these formalisms.
As stated in Sec.~\ref{subsec:statistical}, the neutron capture channel is modeled using 10 effective channels to represent the large number of final states arising from the high level density of the compound nucleus. This treatment is more realistic than using a single channel, as it accounts for spin and parity conservation in the $\gamma$-decay process.

\begin{figure}[H]
    \centering
    \includegraphics[width=\linewidth]{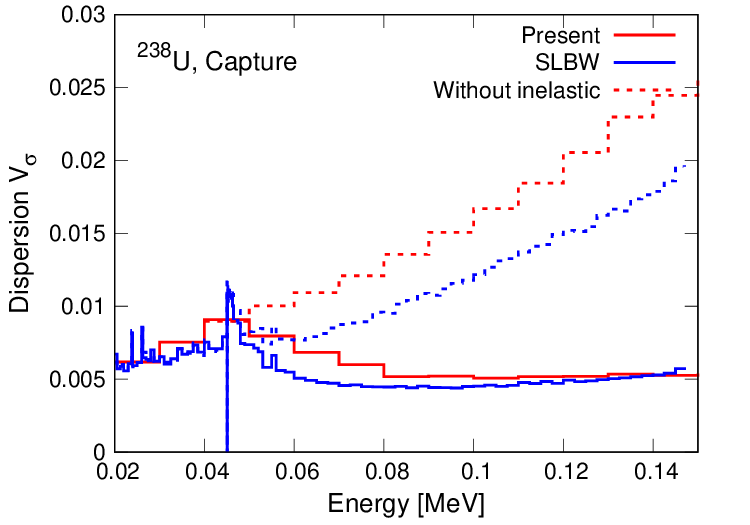}
    \caption{The dispersions $V_{\sigma}$ of the neutron capture cross section distributions. The red dashed line represents the one without inelastic channels.}
    \label{fig:dispersion_capture}
\end{figure}

Figure~\ref{fig:dispersion_capture} presents $V_{\sigma}$ for the neutron capture cross section as a function of the incident neutron energy.
While $V_{\sigma}$ for the total cross section increases monotonically, as shown in Figure~\ref{fig:dispersion_total}, $V_{\sigma}$ for the neutron capture cross section exhibits a peak structure around 0.045 MeV in both the GOE-$S$-matrix and SLBW results.
In the case of $^{238}$U, the first 2$^+$ excited state lies at 0.044916 MeV~\cite{Capote2009}, which corresponds to the threshold at which the inelastic scattering channel opens.
The dashed lines in Fig.~\ref{fig:dispersion_capture} represent $V_{\sigma}$ without the inelastic scattering channel.
The results without the inelastic channel show a monotonically increasing trend, a similar behavior as for the total cross section.
This indicates that the peak observed in $V_{\sigma}$ is caused by the competition between the inelastic scattering and neutron capture reactions.

The URR region stops at 0.15 MeV in the case of $^{238}$U, hence only one inelastic scattering to the 0.045 keV level is involved.
When multiple levels are energetically possible, SLBW cannot include these channels independently but they will be lumped as in Eq.~(\ref{eq:tot_width}).
This would cause some problems to understand the statistical properties of nuclear reaction, when several levels can be excited by incident neutrons. 
The GOE-$S$-matrix model does not have this deficiency, because there is no limitation in the number of inelastic scattering channels as far as the channel transmission coefficients are provided.

%% file: conclusion.tex
We investigated the statistical properties of neutron-induced nuclear reactions using the GOE-$S$-matrix model, which incorporates the Gaussian Orthogonal Ensemble (GOE) into the scattering ($S$) matrix.
This model does not require the average level spacing $D$ and the average decay width $\braket{\Gamma}$ for calculating the $S$ matrix; instead, it employs transmission coefficients as used in the Hauser-Feshbach theory.
We introduced the phase shift of incoming and outgoing channels to remap the GOE eigenchannel energy scale onto the actual laboratory energy.
We calculated the distribution of decay widths, the fluctuating and average cross sections, and the cross section distribution in the unresolved resonance region (URR), which lies between the resolved resonance region (RRR) and the higher energy region. 
Using $^{238}$U as the target nucleus, we adopted the transmission coefficients calculated by CoH$_3$ that reproduce the experimental average cross section.

The distribution of the decay width was calculated by using the $K$-matrix representation and the pole expansion of the $S$ matrix without assuming any statistical distributions.
For the elastic scattering channel, the resulting distribution resembled a $\chi$-squared distribution with degrees of freedom slightly greater than 1, which is consistent with existing theoretical prescriptions~\cite{Moldauer1980,Kawano2015}.
In the case of the neutron capture channel, the distribution approached a Gaussian distribution with a mean of 1, which corresponds to a $\chi$-squared distribution with larger degrees of freedom as expected from the central limit theorem.
The calculated fluctuating and average cross sections were compared with those reconstructed by the single-level Breit-Wigner formula (SLBW) using average resonance parameters taken from ENDF/B-VIII.0.
The distribution of the total cross section was consistent with that obtained from the Reich-Moore formalism, indicating that a smooth transition of the cross section from the RRR to URR was attained.
The peak position of the distribution and the energy dependence of its variance, as calculated by the GOE-$S$-matrix model, showed the same trend as SLBW; however, the model calculation reflected effects arising from the unitarity of the $S$ matrix and resonance interference.
Regarding the distribution of the neutron capture cross section, the calculated distribution resembled that obtained from the Reich-Moore formalism, although differences were observed in the tail region due to the treatment of the neutron capture channel.
Since the Reich-Moore resonances are experimentally observed, we demonstrated that the actual statistical properties can be reproduced by the GOE-$S$-matrix model, where the transmission coefficients employed in the model ensure the Hauser-Feshbach calculations. 
We conclude that the GOE-$S$-matrix model provides a natural and smooth transition between the RRR and the higher energy region, as it employs common model inputs such as transmission coefficients.